\documentclass[11pt, letter]{article}

\usepackage{stolenstyle}

\usepackage{amsmath,epsf,amssymb,latexsym,amsthm,setspace,bbm,array,pifont,enumerate}

\usepackage[matrix,arrow,frame,import,curve,color]{xy}

\DeclareFontFamily{U}{rsf}{}
\DeclareFontShape{U}{rsf}{m}{n}{
  <5> <6> rsfs5 <7> <8> <9> rsfs7 <10-> rsfs10}{}
\DeclareMathAlphabet\Scr{U}{rsf}{m}{n}

\def\CO#1#2{{[#1,#2]}}
\def\AC#1#2{{\{#1,#2\}}}

\def\R{{\mathbb R}}

\def\Img{\operatorname{Im}}

\def\Rea{\operatorname{Re}}

\def\im{\operatorname{im}}

\def\Spin{\operatorname{Spin}}
\def\GG{\operatorname{G}}

\def\so{\operatorname{\mathfrak{so}}}

\def\su{\operatorname{\mathfrak{su}}}
\def\Lu{\operatorname{\mathfrak{u}}}

\def\Lg{\operatorname{\mathfrak{g}}}

\def\p{\partial}
\def\pb{\bar{\partial}}

\def\la{\langle}
\def\ra{\rangle}

\def\ff#1#2{{\textstyle\frac{#1}{#2}}}

\def\cA{{\cal A}}

\def\cG{{\cal G}}

\def\cJ{{\cal J}}
\def\cK{{\cal K}}
\def\cL{{\cal L}}

\def\cO{{\cal O}}

\def\cU{{\cal U}}

\newcommand\zetab{\overline{\zeta}}

\newcommand\thetab{\overline{\theta}}

\newcommand\Sigmab{\overline{\Sigma}}

\newcommand\cb{\overline{c}}

\newcommand\hb{\overline{h}}
\newcommand\ib{\overline{\imath}}

\newcommand\qb{\overline{q}}
\newcommand\rb{\overline{r}}
\newcommand\sbr{\overline{s}}  

\newcommand\zb{\overline{z}}

\newcommand\Cb{\overline{C}}

\newcommand\Gb{\overline{G}}

\newcommand\Jb{\overline{J}}
\newcommand\Kb{\overline{K}}

\newcommand\Qb{\overline{Q}}

\newcommand\Tb{\overline{T}}

\def\cAb{{\overline{\cA}}}
\def\cGb{{\overline{\cG}}}
\def\cJb{{\overline{\cJ}}}
\def\cLb{{\overline{\cL}}}

\def\bz{{{\boldsymbol{z}}}}
\def\bGb{{{{\Gb}}}}
\def\bJb{{{{\Jb}}}}
\def\bqb{{{{\qb}}}}
\def\bQ{{{{Q}}}}
\def\bQb{{{{\Qb}}}}
\def\bT{{{{T}}}}
\def\bTb{{{{\Tb}}}}

\def\tIR{{{\text{IR}}}}

\def\preprint{ ~~ }

\title{Relevant deformations and c-extremization}
\author {Ilarion V.~Melnikov}
\affiliation{Department of Physics and Astronomy,\\
James Madison University, Harrisonburg, VA 22807, USA}
\emailAdd{melnikix@jmu.edu}

\abstract{We consider RG flows obtained by a relevant deformation from  unitary and compact two--dimensional (0,2) SCFTs.  We point out that an N=2 super-Kac-Moody algebra present in the UV is preserved by the flow and does not mix with the R-current.  On the other hand, a direct sum of N=2 algebras in the UV theory leads to a few complications in determining the IR R-symmetry; nevertheless,  in flows without accidental IR symmetries, we determine the IR R-symmetry and show that it maximizes the IR central charge. }

\begin{document}

\maketitle

\section{Introduction}\label{s:intro}

A priori constraints on renormalization group (RG) flows provide key insights into the structure of quantum field theory.  These constraints are often easier to derive in supersymmetric theories, particularly in even dimensions, where they are often related to robust quantities like anomalies in global symmetries.

This note is a comment on one such constraint obtained in~\cite{Benini:2012cz}.  Consider a two-dimensional RG flow that preserves (0,2) supersymmetry and leads to a unitary compact superconformal field theory in the IR.\footnote{A compact CFT has a finite number of states with dimension less than or equal to any $\Delta \in \R$.}  Suppose that the IR R-symmetry arises as a linear combination of symmetries preserved along the RG-flow.  It is shown in~\cite{Benini:2012cz} that the linear combination is determined by finding the extremum of a quadratic trial function $\Cb$; moreover, the extremum value of $\Cb$ is the right-moving central charge of the IR theory, $\cb_{\tIR}$.  

Extremization has the following significance: the trial function is maximized in directions that correspond to left-moving (in our conventions holomorphic) symmetries of the IR theory, and it is minimized in directions that correspond to right-moving (anti-holomorphic) non-R symmetries of the IR theory.  

We point out a simplification for (0,2)-preserving RG flows obtained by relevant deformations of a unitary compact CFT, where the UV N=2 superconformal algebra is a direct sum of decoupled N=2 algebras.  The N=2 algebras that are preserved by the relevant deformation remain decoupled and show up in the IR, each with its own R-symmetry.  It then remains to find the IR R-symmetry in the sector coupled by the deformations.  We show that if the IR R-symmetry is a linear combination of symmetries preserved along the flow, then it is a linear combination of left-moving symmetries and the diagonal R-symmetry of the UV theory.  The exact linear combination is then determined by \textit{maximizing} $\Cb$.

\section{Conserved currents in (0,2) CFT}\label{s:conserved}
Our starting point for constructing the RG flow is a unitary compact SCFT with (0,2) superconformal invariance.  Such a theory will in general have a number of conserved currents that generate a reductive Lie algebra $\Lg_{\text{tot}}$, which consists of a semi-simple component and an abelian component $\Lu(1)^{\oplus r_{\text{tot}}}$.  We will focus on this abelian component in what follows.

Unitarity and compactness guarantee that $\Lu(1)^{\oplus r_{\text{tot}}} = \Lu(1)^{\oplus r} \oplus \Lu(1)^{\oplus \rb}$, and the conserved currents satisfy
\begin{align}
\label{eq:current}
\pb J_\alpha &= 0~,\qquad \alpha =1,\ldots, r~,\nonumber\\
\p \Jb_{\dot\alpha} & = 0~,\qquad \dot\alpha = 1,\ldots, \rb~,
\end{align}
where $\pb = \p/\p\zb$ and $\p = \p/\p z$.\footnote{The careful reader will note that we do note consider the possibility that $r=0$, i.e. there are no left-moving KM symmetries.  As we will see shortly, with our assumptions it is necessary to have at least $r=1$ to obtain a supersymmetric RG flow where the IR R-symmetry is a linear combination of the UV symmetries.}  For $z\neq 0$ the non-vanishing current--current correlation functions are
\begin{align}
z^2\la J_\alpha (z) J_\beta(0) \ra &= K_{\alpha\beta}~,&
\zb^2 \la J_{\dot\alpha} (\zb) J_{\dot\beta}(0) \ra &= \Kb_{\dot\alpha\dot\beta}~,&
z\zb \la J_\alpha(z) \Jb_{\dot\beta} (0)\ra & =0~,
\end{align}
where $K$ and $\Kb$ are symmetric positive matrices.  It will be convenient to normalize the holomorphic currents so that $K_{\alpha\beta}  = \delta_{\alpha\beta}$.

We have yet to use the assumption of N=2 invariance:  we have a superconformal algebra $\cA_{\text{Vir}} \oplus\cAb_{N=2~ \text{sVir}}$, which means that we can organize all of the currents  into supersymmetry multiplets.  

\subsection{Superspace and representations for currents}
It is convenient to describe these multiplets in terms of a (0,2) superspace, with $\bz$ a short-hand for $(z;\zb,\theta^+,\theta^-)$.  
The superspace coordinates $\theta^\pm$ are labeled by their R-charge, and the global superconformal algebra has the representation
\begin{align}
\cJb_0 &= \theta^+ \p_{\theta^+} - \theta^- \p_{\theta^-}~,&
\cLb_0 & = -\zb\pb -\ff{1}{2} (\theta^+\p_{\theta^+} +\theta^-\p_{\theta^-} )~,& \cGb^\pm_{-1/2} & = \p_{\theta^\mp} - \theta^\pm \pb~,& \nonumber\\
\cLb_{-1} & = -\pb~,&
\cLb_{1} & = -\zb^2 \pb~,&
\cGb^\pm_{1/2} & = (\zb-\mp \theta^+\thetab^-) \cG^\pm_{-1/2}~, \nonumber\\
\cL_0 & = -z\p ~,&
\cL_{-1} & = -\p~,&
\cL_{1} & = -z^2\p~.
\end{align}
Here $\p = \p/\p z$, $\pb = \p/\p\zb$, and $\p_{\theta^\pm} = \p / \p \theta^\pm$.

The holomorphic currents $J_\alpha(z)$ are easily described:  they are primary operators of weight $(h,\hb) = (1,0)$ and therefore are in the trivial representation of $\cAb$.  The same holds for the holomorphic stress tensor $\bT(z)$.

On the other hand, the anti-holomorphic currents reside in non-trivial representations of $\cAb$.  Perhaps the most familiar example of this is offered by the superconformal current multiplet of $\cAb$:
\begin{align}
\Sigmab(\bz) = \bJb(\zb) + \theta^+ \bGb^-(\zb) -\theta^-\bGb^+(\zb) + 2\theta^+\theta^- \bTb(\zb)~.
\end{align}
$\bJb$ is the R-current, the $\bGb^\pm$ are the supercurrents, and $\bTb$ is the right-moving stress tensor.  In particular, the supersymmetry charges are the modes $\bGb^\pm_{-1/2}$.

If $\bGb^\pm$ are the only conserved spin $3/2$ currents in the theory, then $\Sigmab$ is the unique multiplet with a (0,1) current as the lowest component.  A current $\Jb_{\dot\alpha}$ cannot appear as the top component of a multiplet, because then the lowest component would necessary have scaling dimension $0$.  In a compact CFT the only such operator is the identity.  It is possible for $\Jb_{\dot\alpha}$ to appear as a middle component of an N=2 multiplet.  Once we restrict to abelian currents, the resulting multiplet must necessarily be a short multiplet, either chiral or anti-chiral, because otherwise there would be a (0,1) current charged with respect to the R-symmetry.

Thus, when $\bGb^\pm$ currents are the unique spin $3/2$ conserved currents, the remaining abelian (0,1) currents assemble into chiral/antichiral N=2 super Kac-Moody (SKM) multiplets of the form
\begin{align*}
\Psi_i &= \psi_i +\sqrt{2} \theta^+ \Jb_i + \theta^+\theta^- \pb \psi_i~,\nonumber\\
\Psi^\dag_{\ib} &= \psi^\dag_{\ib} +\sqrt{2}\theta^- \Jb^\dag_{\ib} -\theta^+\theta^- \pb \psi_{\ib}^\dag~,
\end{align*}
where the $\psi_i$ and $\psi^\dag_{\ib}$ are anti-holomorphic operators with $(h,\hb) = (0,1/2)$, R-charge respectively $+1$ and $-1$, and non-vanishing two-point functions 
\begin{align*}
\zb \la \psi^\dag_{\ib} (\zb) \psi_i (0) \ra = \cK_{\ib i}~,
\end{align*}
where $K$ is a positive Hermitian matrix.  Each $\Jb_i$ is a complex combination of two of the $\Jb_{\dot\alpha}$, and the non-vanishing two-point functions are
\begin{align*}
\zb^2 \la \Jb^\dag_{\ib}(\zb) \Jb_i(0)\ra = \cK_{\ib i}~.
\end{align*}
The $\cK_{\ib i}$ then determine the $\Kb_{\dot\alpha\dot\beta}$ in~(\ref{eq:current}).
Suppose we have $i=1,\ldots, d$ such multiplets.  If $d=1$, then we find three abelian right-moving currents: $\Rea\Jb$, $\Img \Jb$ , and $:\!\psi\psi^\dag\!\!\!:$~.  For $d>1$ the free fermions generate an $\so(2d)$ level $1$ KM algebra.  The N=2 SKM has a Sugawara construction~\cite{Sevrin:1988ps,Kazama:1988uz}.  For instance, for $d=1$ we obtain the free-field representation familiar from toroidal compactification:
\begin{align*}
\bJb &= :\psi\psi^\dag:~,&
\bGb^+& = \sqrt{2} \psi \Jb^\dag~,&
\bGb^- & = \sqrt{2} \psi^\dag \Jb~,&
\bTb & =~ : \Jb\Jb^\dag: -\ff{1}{2} ( : \psi^\dag\pb\psi: + : \psi \pb \psi^\dag : )~.
\end{align*}
This N=2 algebra algebra has central charge $\cb = 3$.

\subsection{A direct sum of N=2 algebras}

There is another possibility for the appearance of additional (0,1) abelian currents.  Suppose $\cAb$ decomposes into a direct sum of N=2 algebras: $\cAb = \oplus_{\dot\alpha=1}^{\rb} \cAb_{\dot\alpha}$.  In this case we have $\rb$ supercurrent multiplets
\begin{align}
\Sigmab_{\dot\alpha}(\bz) = \bJb_{\dot\alpha}(\zb) + \theta^+ \bGb_{\dot\alpha}^-(\zb) -\theta^-\bGb_{\dot\alpha}^+(\zb) + 2\theta^+\theta^- \bTb_{\dot\alpha}(\zb)~.
\end{align}
The OPE $\Sigma_{\dot\alpha} (\bz_1) \Sigma_{\dot\beta}(\bz_2)$ is regular unless $\dot\alpha=\dot\beta$, and the sum $\Sigmab = \sum_{\dot \alpha} \Sigmab_{\dot\alpha}$ generates the diagonal N=2 algebra; the top component of $\Sigmab$ is the energy momentum tensor of the SCFT, i.e. the operator that couples to a background world-sheet metric.   The central terms in each of the algebras are fixed by superconformal invariance and the two-point function of the lowest components:
\begin{align}
\label{eq:JJ}
\zetab_{12}^2 \la \Sigmab_{\dot\alpha}(\bz_1) \Sigmab_{\dot\beta} (\bz_2) \ra = \frac{\cb_{\dot\alpha}}{3} \delta_{\dot\alpha\dot\beta}~\qquad\qquad \zetab_{12} = \zb_{12} -\theta_1^+\theta_2^- -\theta_1^-\theta^2_+~.
\end{align}

More generally, the supercurrent multiplet may be reducible but not necessarily decomposable.\footnote{This terminology is familiar in the context of supercurrent multiplets from~\cite{Dumitrescu:2011iu}:  a multiplet is reducible if it is a sum of two separate supersymmetry multiplets; a multiplet is decomposable if it can be split into two decoupled supercurrent multiplets.} In other words, $\Sigmab =\sum_{\dot\alpha=1}^{\rb} \Sigmab_{\dot\alpha}$, where the $\Sigmab_{\dot\alpha}$ are N=2 quasi-primary but not necessarily commuting operators.
Given a reducible supercurrent multiplet, the necessary and sufficient conditions for decomposability into $\rb$ components are as follows:
\begin{enumerate}
\item the OPE of the $\Sigmab_{\dot\alpha}$ closes;
\item the $\bJb_{\dot\alpha}$ are abelian currents with two-point function~(\ref{eq:JJ});
\item the $\bGb^\pm_{\dot\alpha}$ carry charges $\pm \delta_{\dot\alpha\dot\beta}$ with respect to $\bJb_{\dot\beta}$;
\item the $\bTb_{\dot\alpha}$ are $\bJb_{\dot\beta}$--neutral.
\end{enumerate}
The first condition implies that the algebra is determined by the two- and three-point functions of the $\Sigmab_{\dot\alpha}$.  The former are fixed by the second condition and superconformal invariance, while the latter satisfy \begin{align}
\la \bJb_{\dot\alpha}(\zb_1)\bJb_{\dot\beta}(\zb_2)\bJb_{\dot\gamma}(\zb_3)\ra & = 0~,&
\la \bJb_{\dot\alpha}(\zb_1)\bTb_{\dot\beta}(\zb_2)\bTb_{\dot\gamma}(\zb_3)\ra & = 0~, \nonumber\\
\la \bJb_{\dot\alpha}(\zb_1) \bGb^+_{\dot\beta}(\zb_2)\bGb^-_{\dot\gamma}(\zb_3) \ra & =
\frac{2}{3}\cb_{\dot\alpha} \frac{ \delta_{\dot\alpha\dot\beta} \delta_{\dot\beta\dot\gamma}}{\zb_{12}\zb_{13} \zb_{23}^2}~.
\end{align}
The supersymmetry relations
\begin{align}
\bGb^\pm_{\dot\alpha} & = \CO{\bGb^\pm_{-1/2}}{\bJb_{\dot\alpha}(\zb)}~,&
\AC{\bGb^\pm_{-1/2}}{\bGb^\mp_{\dot\alpha}(\zb)} & = \pm 2 \bTb_{\dot\alpha}(\zb) + \pb \bJb_{\dot\alpha} (\zb)
\end{align}
determine the remaining three-point functions
\begin{align}
\la \bJb_{\dot\alpha}(\zb_1)\bJb_{\dot\beta}(\zb_2)\bTb_{\dot\gamma}(\zb_3)\ra~, &&
\la \bGb^+_{\dot\alpha}(\zb_1)\bGb^-_{\dot\beta}(\zb_2)\bTb_{\dot\gamma}(\zb_3)\ra~, &&
\la \bTb_{\dot\alpha}(\zb_1)\bTb_{\dot\beta}(\zb_2)\bTb_{\dot\gamma}(\zb_3)\ra~. 
\end{align}
For instance, we have
\begin{align}
2 \la  \bJb_{\dot\alpha}(\zb_1)\bJb_{\dot\beta}(\zb_2)\bTb_{\dot\gamma}(\zb_3)\ra & =
2 \la  \bJb_{\dot\alpha}(\zb_1)\bJb_{\dot\beta}(\zb_2)\bTb_{\dot\gamma}(\zb_3)\ra +
\pb_3 \la  \bJb_{\dot\alpha}(\zb_1)\bJb_{\dot\beta}(\zb_2)\bJb_{\dot\gamma}(\zb_3)\ra \nonumber\\[2mm]
&=
 \la \bJb_{\dot\alpha}(\zb_1)\bJb_{\dot\beta}(\zb_2)\AC{\bGb^+_{-1/2} }{\bGb^-(\zb_3)} \ra \nonumber\\[2mm]
 &=
 \la \bGb^+_{\dot\alpha}(\zb_1) \bJb_{\dot\beta}(\zb_2) \bGb^-_{\dot\gamma}(\zb_3)\ra
 +\la\bJb_{\dot\beta}(\zb_1) \bGb^+_{\dot\alpha}(\zb_2)  \bGb^-_{\dot\gamma}(\zb_3)\ra~ \nonumber\\[2mm]
 & = \frac{2}{3} \cb_{\dot\alpha} \frac{\delta_{\dot\alpha\dot\beta} \delta_{\dot\beta\dot\gamma}}{\zb_{12}^2\zb_{23}^2}~.
\end{align}
The remaining correlators are determined by very similar manipulations.  So, although in general the three-point functions of superconformal descendants are not fixed in terms of those of the N=2 primaries, three-point functions of conserved currents can be determined completely by using the current algebra and superconformal invariance.\footnote{For more details on the uses of superconformal Ward identities in (0,2) theories the reader might consult \cite{West:1990tg}; a four-dimensional example of current $3$-point functions in a superconformal theory can be found in~\cite{Fortin:2011nq}.}

Are there reducible but indecomposable supercurrent multiplets?  There is one obvious example:  the small N=4 algebra.  In this case the charge $+2$ $\su(2)$ current $J^{++}$ is the lowest component of a chiral multiplet, where the fermi component is the additional supercharge $\bGb'^+$.  Similarly, $J^{--}$ and $\bGb'^-$ reside in an anti-chiral multiplet of the diagonal N=2 algebra.  More generally, two-dimensional superconformal algebras are classified under the assumptions that conserved currents have spins in $\{0,1/2,1,3/2,2\}$, a unique energy momentum tensor, and their OPE closes~\cite{Ramond:1976qq,Sevrin:1988ew,Spindel:1988sr,Sevrin:1988ps}; a summary may be found in~\cite{Polchinski:1998rr}.   Some indecomposable examples of N=1 supercurrent multiplets arise in $\GG_2$  and $\Spin(7)$ holonomy sigma models~\cite{Shatashvili:1994zw,FigueroaOFarrill:1996hm}.  

It would be interesting to describe other reducible but indecomposable supercurrent multiplets of the N=2 algebra, but in this work we will restrict attention to UV fixed points that realize a sum of irreducible $N=2$ algebras.  

We point out that a supercurrent multiplet $\Sigmab$ with $\rb>1$ components is a counter-example to the assertion that the R-current (i.e. the lowest component of $\Sigmab$ ) has vanishing anomaly with any other right-moving current.

\section{Relevant deformations }\label{s:relevant}
Having reviewed the structure of symmetry currents in a compact unitary (0,2) SCFT, we now consider (0,2) supersymmetric relevant deformations of the theory.   In this section we will show that N=2 SKM multiplets do not participate in the RG flow, and we will also describe the symmetries of the UV theory preserved by a relevant deformation.

As was shown in~\cite{Bertolini:2014ela}, supersymmetric relevant deformations are in one to one correspondence with chiral primary operators $\cU_I$ with weights $h_I = \bqb_I/2 +1/2$ and $\hb_I =\bqb_I /2$.\footnote{By a supersymmetric deformation we mean one that preserves the full (0,2) supersymmetry.}  Here $\bqb_I <1$ is the R-charge of $\cU_I$ with respect to the diagonal R-symmetry $\bJb$.  When there are multiple N=2 supercurrent multiplets, i.e. $\rb >1$, we can write $\bqb_I = \sum_{\dot\alpha}\bqb_I^{\dot\alpha}$~, where $\bqb_I^{\dot\alpha}$ is the R-charge of $\cU_I$ with respect to the current $\bJb_{\dot\alpha}$.  The operator $\cU_I$ is chiral primary with respect to the diagonal superconformal algebra if and only if it is chiral primary with respect to each of the simple sub-algebras.  Hence, we can assume $\bqb_I^{\dot\alpha} \ge 0$.

We now argue that supersymmetric relevant deformations are neutral under all right-moving currents in N=2 SKM multiplets.  The only non-irrelevant deformations that involve the degrees of an N=2 SKM multiplet are of the form $K(z)\psi(\zb)$, where $K(z)$ is a left-moving (1,0) current, and $\psi(\zb)$ is the lowest component of a chiral N=2 SKM multiplet.  Every such operator is marginal at leading order in conformal perturbation theory. This point was already made in~\cite{Bertolini:2014ela}, but we repeat it here with new emphasis:  a supersymmetric relevant deformation leaves every N=2 SKM unbroken.  These remain symmetries along the flow and of course also at the IR fixed point. Thus, we can and will ignore the N=2 SKM sector in our search for the IR R-current.

Let us determine the symmetries preserved by a supersymmetric relevant deformation that involves some set of operators $\cU_I$, with $I=1,\ldots, N$.  For this we just need to consider the form of the deformation and classify the currents that remain conserved in the presence of the deformation.   The deformation action has the form
\begin{align}
 \sum_{I=1}^N \lambda^I \int d^2 z ~\AC{ \bGb^-_{-1/2}}{\cU_I} + \text{h.c.}~ 
& = \sum_{I=1}^N\sum_{\dot\beta=1}^{\rb} \lambda^I \int d^2z~~ \cO_{I\dot\beta}+ \text{h.c.}~,
\end{align}
where $\cO_{I\dot\beta} = \AC{(\bGb^-_{\dot\beta})_{-1/2}}{\cU_I}$, and the $\lambda^I$ are coupling constants.  The operator $\cO_{I\dot\beta}$ is non-zero if and only if $\bqb^{\dot\beta}_I >0$.  

Denote the conserved  charges corresponding to $J_\alpha$ and $\bJb_{\dot\alpha}$ of the undeformed theory by, respectively, $\bQ^\alpha$ and $\bQb^{\dot\alpha}$, and consider a general combination of these
\begin{align*}
Q[s] = \sum_{\alpha} s_\alpha Q^\alpha +  \sum_{\dot\alpha} \sbr_{\dot\alpha} \bQb^{\dot\alpha}~.
\end{align*}
A non-zero operator $\cO_{I\dot\beta}$  satisfies
\begin{align}
-i\CO{Q^\alpha}{\cO_{I\dot\beta}} &= q^\alpha_I\cO_{I\dot\beta} ~,&
-i\CO{\bQb^{\dot\alpha}}{\cO_{I\dot\beta}} & = \delta_{\dot\alpha\dot\beta} (\bqb_I^{\dot\alpha}-1) + (1-\delta_{\dot\alpha\dot\beta}) \bqb^{\dot\alpha}_I~
\end{align}
and is therefore neutral with respect to $Q[s]$ if and only if 
\begin{align}
\label{eq:symreq}
\sbr_{\dot\beta}  = \sum_{\alpha=1}^r q^\alpha_I s_\alpha  +\sum_{\dot\alpha=1}^{\rb} \bqb^{\dot\alpha}_I \sbr_{\dot\alpha}~.
\end{align}
When this holds, it is easy to show that to leading order in conformal perturbation theory
\begin{align}
\pb \sum_{\alpha=1}^r s_\alpha J_\alpha + \p \sum_{\dot\alpha=1}^{\rb} \sbr_{\dot\alpha} \bJb_{\dot\alpha} = 0~,
\end{align}
and we will assume that this current remains conserved along the RG flow.

\section{Extremization and the IR R-symmetry}\label{s:extreme}

We now assume that the entire RG flow (as opposed to just the infinitesimal deformation) is supersymmetric and leads to a compact unitary CFT in the IR.  Furthermore, we assume that the IR R-symmetry arises as a linear combination of the symmetries  preserved along the flow. We will show that the particular linear combination is determined by superconformal invariance.  The main tool is the same as in~\cite{Benini:2012cz}, i.e. `t Hooft anomaly matching.\footnote{Some earlier applications to similar questions in the (0,2) context  were made in~\cite{Distler:1995mi,Tong:2008qd,Melnikov:2009nh}.}

Observe that if the deformation is invariant under a $\cAb_{\dot\beta}$ sub-algebra, then $\cAb_{\dot\beta}$ is preserved along the RG flow and will show up as a summand in the IR superconformal algebra.  It will not mix with the superconformal algebra of the ``interacting'' part of the theory.  As we showed above, any N=2 SKM must be such a decoupled summand, but there may be other decoupled factors as well.  There is no mystery about the R-symmetry for each of the decoupled factors:  it remains exactly the same and never mixes with the ``interacting'' sector with non-trivial RG flow.  So, we can now turn to the remaining R-symmetry question:  how do identify the R-symmetry in the interacting sector?

To keep the notation simple we will use the same $\dot \alpha$ index to refer just to the ``interacting sub-algebras;''  i.e. for every $\dot \alpha$ there is some $I$ such that $\cO_{I\dot\alpha} \neq 0$.   With that simplification in hand, taking a look at~(\ref{eq:symreq}), we conclude $\sbr_{\dot\beta} = \sbr_0$, a constant independent of $\dot\beta$.  Thus, we can simplify~(\ref{eq:symreq}) to
\begin{align}
\label{eq:symreq2}
\sum_{\alpha=1}^r q_I^\alpha s_\alpha & = \sbr_0 (1-\qb_I)\qquad\qquad\text{for all $I$}~.
\end{align}
When this is satisfied, we have a conserved charge
\begin{align}
\label{eq:cons}
Q[s] = \sum_{\alpha=1}^r s_\alpha Q^\alpha + \sbr_0 \bQb~,
\end{align}
where $\bQb = \sum_{\dot\alpha=1}^{\rb} \bQb^{\dot\alpha}$ is the diagonal R-charge of the unperturbed theory.  We wish the $Q[s]$ to be an R-symmetry along the flow, which requires $\sbr_0 = 1$.\footnote{This follows because we want to assign R-charges $\pm 1$ to $\theta^\pm$, and $\sbr_0=1$ is the correct choice.}  So, packaging the $q_I^\alpha$ into an $N\times r$ matrix $L$, $(L)_I^\alpha = q_I^\alpha$, we now recast~(\ref{eq:symreq2}) as
\begin{align}
\label{eq:reqsymfin}
L s & = \rho~,
\end{align}
where $\rho_I = 1-\qb_I$.  Therefore a succinct form for the necessary and sufficient condition for the flow to preserve an R-symmetry is
\begin{align}
 \rho \in \im L~.
\end{align}
We now see that, as promised, a (0,2) supersymmetric deformation of  a unitary compact SCFT is only possible if $r >0$; otherwise~(\ref{eq:reqsymfin}) implies $ \qb_I = 1$ for all $I$, i.e. the deformation is marginal.  When it exists, the solution for $s$ is ambiguous if $\dim\ker L= n>0$.  Fix an orthonormal (with respect to the standard Euclidean metric on $\R^r$) basis $\{\omega_1, \omega_2, \ldots,\omega_n\}$ for $\ker L$, i.e. $\omega^T_i \omega_j =\delta_{ij}$.\footnote{We are using the simplification that the left-moving currents of the UV theory were normalized with $K_{\alpha\beta} = \delta_{\alpha\beta}$.}  Given any solution to~(\ref{eq:reqsymfin}), say $s= \sigma$, we can form a trial solution
\begin{align}
s(t) = \sigma+ \sum_{i=1}^n t^i \omega_i~.
\end{align}
Our goal is to determine the $n$ parameters $t^i$ such that $\bQb[s(t)]$ is the IR R-symmetry.  This is easy once we understand the physical significance of the $t^i$ in terms of the structure of the conserved currents.  For each $\omega_i$ we obtain in the IR a left-moving KM algebra, while choosing $s_\alpha = s_\alpha(t)$ in~(\ref{eq:cons}) will yield the IR R-symmetry.  In the IR CFT the left-moving KM symmetries have no mixed anomalies with right-moving KM symmetries.  However, since the anomalies are RG-invariant, we can also compute them in the UV theory in terms of the current-current two-point functions.  There are two interesting classes of these for us:
\begin{align}
\frac{1}{3}\Cb(t)&= \zb^2 \la \bJb(z) \bJb(0)\ra - z^2 \la J_{\text{trial}}(z) J_{\text{trial}}(0)\ra~, \nonumber\\
X_i & = z^2 \la J_i(z) J_{\text{trial}}(0)\ra~,
\end{align}
where
\begin{align}
J_{\text{trial}} & = \sum_{\alpha=1}^r s_\alpha(t) J_\alpha~,&
\Jb_{\text{trial}} & = \bJb~,\nonumber\\
J_i & = \sum_{\alpha=1}^r (\omega_i)_\alpha J_\alpha~.
\end{align}
To ensure that the trial R-symmetry has no mixed anomalies with the remaining left-moving symmetries we must choose the parameters $t^i$ such that $X_i = 0$.  With our chosen normalization $z^2\la J_\alpha(z) J_\beta(0) \ra = \delta_{\alpha\beta}$, this determines 
\begin{align}
t_i  & = -\omega^T_i \sigma~,&
s(t) & =  \sigma_{\perp} = \sigma - \sum_i (\omega_i^T \sigma)\omega_i~.
\end{align}
Once this is satisfied, the IR R-symmetry is determined.  The IR central charge is given by evaluating $\Cb(t)$:
\begin{align}
\cb_{\tIR} &= \cb_{\text{UV}} - 3\sigma^T_{\perp} \sigma_{\perp}~.
\end{align}
The IR R-symmetry \textit{maximizes} the trial function $\Cb(t)$, and the maximum value is $\cb_{\tIR}$. 

\section{Conclusions}
We have shown that with our assumptions there is a simplification in the c-extremization of~\cite{Benini:2012cz}, and the two-dimensional R-symmetry is determined much as in N=1 d=4 flows, which maximize a trial function for the central charge $a$~\cite{Intriligator:2003jj}. 
This observation was inspired by~\cite{Bertolini:2014ela}, where the c-extremization result of~\cite{Benini:2012cz} was used to study basins of attraction in (0,2) Landau-Ginzburg (LG) theories.  In these asymptotically free RG flows we observed a number of important features.  First, we noted empirically that in theories where the quasi-homogeneous superpotential had an isolated minimum, and thus a normalizable ground state, there were no examples of right-moving non-R symmetries in N=2 SKM multiplets.  Indeed, whenever the UV theory had an irreducible supercurrent multiplet the quasi-homogeneous superpotential did not admit any symmetries where the mixed anomaly extracted from the two-point function had non-positive eigenvalues.  Any RG flow from an SCFT obtained as an IR fixed point of such a LG theory provides an example of the SCFTs considered in this note.  

The second observation from~\cite{Bertolini:2014ela} that bears on the above results is that accidental symmetries that mix with the R-current are to be found even in these (0,2) simple LG flows.  We expect this to be a typical feature in (0,2) RG flows.  In some cases we expect that unitarity constraints combined with our observation may help to uncover accidental symmetries in non-trivial (0,2) RG flows.

\section*{Acknowledgements}  I thank M.R.~Plesser, A.B.~Royston and the anonymous referee for comments on the manuscript and S.~Sethi for pointing out reference~\cite{FigueroaOFarrill:1996hm}.  

\bibliographystyle{./utphys}
\bibliography{./bigref}

\end{document}